\documentstyle[prl,twocolumn,aps]{revtex}
\begin{document}
\draft
\title{Local electronic properties of single wall nanotube 
circuits measured by conducting-tip AFM}

\author{M. Freitag, M. Radosavljevi\'{c}, W. Clauss, and A. T. Johnson}
\address{Department of Physics and Astronomy, University of Pennsylvania, Philadelphia, PA 19104}

\date{\today}
\maketitle

\begin{abstract}
We use conducting-tip atomic force microscopy (AFM) to measure local electronic properties of single wall carbon nanotube (SWNT) circuits on insulating substrates. When a voltage is applied to the tip and AFM feedback is used to position the tip, images formed from the tip-sample tunnel current have single tube resolution (near 1 nm diameter), more than an order of magnitude better than simultaneously acquired topographic AFM images. By finding points where the tip-sample current is zero, it is possible to measure the electrochemical potential within the circuit, again with nanometer resolution. Such measurements provide compelling evidence that nanotubes within a bundle have only weak electronic coupling. Finally the AFM tip is used as a local electrostatic gate, and the gating action can be correlated with the structure of the SWNT bundle sample. This technique should be useful for a broad range of circuits containing SWNTs and other molecules.

\end{abstract}



Single wall carbon nanotubes (SWNTs) \cite{Iijima,review} are a fascinating class of molecules whose electronic properties are exquisitely sensitive to local structural, mechanical, and electrostatic perturbations. For example, a slight change in the wrapping of the constituent graphene sheet opens an energy gap of 0.6 eV in the band structure of a previously metallic 1.4 nm diameter nanotube \cite{Hamada,Saito,Mintmire}. It was predicted theoretically \cite{Chico} and recently demonstrated in experiment that defects, impurities, and mechanical deformation can lead to a rich variety of functional devices such as rectifying diodes \cite{Antonov,Yao} and double quantum dots \cite{Lefebvre}. Coupled with recent progress in Atomic Force Microscope (AFM) manipulation of SWNTs \cite{Lefebvre} and multi-walled nanotubes \cite{avouris,superfine}, these results imply that it may soon be possible to engineer and fabricate single-nanotube molecular devices. 

Methods to probe the local electronic properties of SWNT circuits are needed to unravel the important design principles. Conventional transport measurements characterize nanotube circuits as a whole, with no local information. Scanning tunnel microscopy (STM) and scanning tunnel spectroscopy (STS), in contrast, give detailed local information about the electronic structure \cite{Wildoer,Odom,Clauss2} of individual nanotubes and SWNT bundles. These have not been applied to nanotube circuits because the tip-sample tunnel current, used for z-position feedback, drops to zero over an insulating substrate, causing the tip to ``crash" into the sample. 

In this work, we marry the strengths of these techniques and use conducting tip Atomic Force Microscopy (CT-AFM) to inspect and make the first measurements of the {\it local} electronic properties of SWNT bundles in circuits with nanometer-scale resolution. In contrast to earlier work \cite{Dai}, we use non-contact AFM to probe delicate SWNT samples that may not tolerate the high forces associated with contact-mode AFM. We make the following observations: First, CT-AFM tunnel current images of SWNTs in circuits can have 1 nm resolution, much better than that of standard AFM. Second, we measure the local voltage within a SWNT bundle and obtain compelling evidence that nanotubes within the bundle are only weakly coupled electrically, as predicted by theory \cite{ahmed}. We also find that the resistance of two electrical contacts to an individual SWNT bundle in a circuit can vary considerably, a possibility that was not considered in earlier experiments. Finally, we demonstrate that the conducting AFM tip can be used as a {\it local gate} of the SWNT circuit, and we correlate the strength of the local gating with the bundle structure. These encouraging first results indicate that tapping mode CT-AFM is likely to play a key role in the development of functional electronic devices based on SWNTs and other molecular species.

Samples for these experiments are SWNT circuits on silicon wafers with a 300 nm oxide layer. SWNTs are made by laser ablation \cite{review,thess}, and the as-grown material dispersed in dimethylformamide (to give small bundles with diameter below 3 nm) or isopropanol (bundle diameter near 10 nm). For some experiments (e.g., Fig.\ 1) Cr/Au electrodes are patterned by electron beam lithography and then drops of SWNT solution spin-coated onto the substrate, dispersing nanotubes that contact the electrodes at random. In other experiments (Figs.\ 2 and 3), SWNTs are first dispersed as above then located by AFM, and electron beam lithography is used to pattern electrodes that connect bundles to large contact pads. 

The AFM we use is based on a vibrating quartz ``needle sensor" (Omicron Instruments) whose mechanical oscillation (amplitude of 1 - 2 nm, tip-sample force of a few nanonewtons \cite{Clauss3}) is detected electrically. Conducting AFM tips (coated with tungsten carbide or doped diamond) are attached with conducting epoxy to the end of the needle sensor using a micromanipulator. All measurements are performed under ambient lab conditions. 

AFM image resolution is typically limited by the tip radius to 10 nm or more, so small features like single tubes in a bundle are not resolved. In contrast, the exponential distance dependence of the tunnel current gives STM of SWNTs much higher (even atomic) resolution \cite{Wildoer,Odom,Clauss2,Clauss1}. STM samples must conduct everywhere, but with CT-AFM, we can apply a voltage between the tip and sample, use the mechanical force signal for feedback, and create an image from the tip-sample tunnel current. With this technique, we have previously achieved near-atomic resolution images of SWNT bundles lying on a gold substrate \cite{Clauss3}.

Here we extend the approach and use this imaging mode to examine a SWNT bundle electrically connected to a single electrode on an otherwise insulating substrate. The upper inset to Fig.\ 1 is the AFM topographic image showing the SWNT bundle and the gold electrode; individual tubes within the bundle are not resolved. The lower inset is the simultaneously obtained current signal with the tip biased at $-2$ V and the gold electrode grounded. As found earlier \cite{Clauss3}, considerably more current flows from the AFM tip to the nanotubes (which are soft and deform under pressure from the tip) than to the harder gold electrode. In the expanded image (tip bias of $-1$ V), single tubes are clearly resolved within the bundle, a resolution enhancement we ascribe to the tunnel current's exponential sensitivity to tip-sample separation. We have found that this enhanced resolution imaging mode is an excellent tool for inspecting nanometer-scale circuits, even on substrates where surface roughness would hinder the detection of nanometer-size objects by conventional AFM. 

With a related approach we measure the voltage {\it inside} a nanotube circuit (Fig.\ 2). Figure 2(a) is a topographic AFM image of a 8 nm high SWNT bundle (blue) connected to two gold electrodes (yellow) on a SiO$_2$ substrate (purple). A potential difference of $0.6$ V exists between the two electrodes, creating a transport current of 1 $\mu$A along the bundle. The AFM tip is biased at an intermediate voltage ($0.45$ V), the mechanical force signal is used as feedback, and we image the tip-sample current (Fig.\ 2(b)). When the local voltage in the circuit is greater than the tip voltage, current flows from the sample to the tip (blue); when the tip voltage is larger, current flows from the tip to the sample (red). Points where the local voltage equals the tip voltage and zero current flows can be located with a spatial resolution of just a few nanometers. Figure 2(b) demonstrates the intriguing fact that regions of different potential coexist along the bundle cross-section. Images with similar structure are frequently obtained in this AFM mode and are completely stable when the scan direction is reversed, so we conclude that this is not an artifact associated with sample topography. Such images provide compelling evidence that current flow is not uniform within the bundle; instead the current carrying path meanders through the bundle (predominantly confined to metallic tubes), and tubes in the bundle are only weakly coupled electronically. This is expected behavior for bundles that consist of different species of SWNTs \cite{ahmed}, as is believed to be the case for samples produced by either laser ablation or carbon arc. Methods to overcome weak coupling between neighboring nanotubes will be needed to realize proposed applications of SWNTs or SWNT bundles as nanoelectronic interconnects. 

In some circuits the contact resistance to a bundle is much larger than the intrinsic bundle resistance, so we measure nearly a constant voltage along the full bundle length of about 1 $\mu$m. We often find that the bundle voltage is almost equal to that of one of the two contacts: we conclude that the contact resistances differ strongly (by as much as a factor of 10), a possibility that has not been considered in earlier experiments. In other circuits, the contact resistance can be so small that voltage changes {\it within} the bundle are resolved, as seen in Fig.\ 2. This technique enables future experiments to create detailed maps of the voltage profile within SWNT circuits for comparison to predictions for transport within these nearly ideal one-dimensional systems. Since the insulating substrate contributes no signal with our approach, it may have significant advantages for this application over related methods that probe local voltages through charge interactions (e.g., electrostatic force microscopy (EFM) \cite{Arnason}), where the substrate dielectric response contributes to the measured image. 

The CT-AFM tip can be used as a local {\it stimulus} for the SWNT circuit, as well as a local probe as described above. In conventional transport measurements, individual semiconducting nanotubes exhibit a strong field effect due to a nearby gate voltage \cite{Antonov,Tans}, as do SWNT bundles where semiconducting tubes carry a large fraction of the current. In contrast, almost no field effect is observed for metallic nanotubes \cite{Yao} or bundles that are heavily doped by exposure to potassium or other elements \cite{roland}. We use the conducting AFM tip as a scanned local gate \cite{Eriksson} of a bundle and find that the gating action is inhomogeneous and correlated with bundle structure. 

For this experiment, the tip-sample force is small, so the tip acts as a capacitively coupled local gate, and no tunnel current flows from the tip to the sample. Figure 3(a) is a topographic AFM image of a bundle (blue) connected to two gold electrodes (yellow). For most of its length, the bundle has a diameter of 8 nm, but an expanded image (Fig.\ 3(b)) shows that the bundle narrows at the point marked by the arrow to just 1.5 nm, the diameter of a single tube. A bias voltage drives current through the circuit which is used as the image signal while the conducting AFM tip (biased at -2 V) is scanned above the sample (Fig.\ 3(c)). A superimposed contour map shows the bundle topography. Over most of the image there is no field effect, indicating that in regions where the bundle has diameter is near 8 nm, current is carried by one or more metallic nanotubes, which are not sensitive to electric field. When the tip is within 30 nm of the narrow region, however, the current increases by nearly a factor of two. The most likely explanation is that in this 1.5 nm diameter region where one or just a few nanotubes contribute to transport, a large fraction or even all of the current is carried by semiconducting SWNTs, which display a ``p-type" response to the gate voltage (i.e., the majority carriers are holes, so a negative gate voltage leads to increased current) \cite{Antonov,Tans}. The spatial resolution is of order 30 nm, which we ascribe to AFM tip curvature and not the much smaller tip-sample separation of 1 - 5 nm. When a positive voltage is applied to the tip (Fig.\ 3(d)), we make a surprising observation: we expect to find only a current decrease, complementary to the image of Fig.\ 3(c), but instead observe a small region with weak ``n-type" behavior (yellow) right next to a p-type region (blue). At this point we can not explain this last finding; more experiments are needed to determine whether this is characteristic of a naturally occurring p-n junction \cite{Chico,kawazoeapl} or some other nanotube structure.

In conclusion, we have developed hybrid scanning probe modes to inspect and characterize SWNT circuits on insulating substrates. The ability to pass a tunnel current between tip and sample in tapping-mode AFM makes it possible to image functioning circuits with 1 nm resolution, an order of magnitude better than is achieved with conventional AFM. The tunnel current also enables the determination of the voltage at specific points with resolution of a few nanometers. Finally, the tip can be used as a local gate of the circuit, and the gating action correlates with sample structure with resolution limited by AFM tip curvature. In the future we plan to apply these techniques to more intricate nanotube circuits, and circuits containing other types of electrically active molecules. 

The high-quality nanotube material used in this work was provided by the Smalley group at Rice University. We acknowledge useful discussions with David Bergeron, James Hone, Charles Kane, Jay Kikkawa, Jacques Lefebvre, and Eugene Mele. This work was supported by the NSF under award number DMR-9802560 and Deutsche Forschungsgemeinschaft (W.C.). A.T.J. recognizes the support of the Packard Foundation and an Alfred P.\ Sloan Research Fellowship.

%
%

\begin{figure}
\caption{(color) High resolution tunnel current image of a SWNT bundle contacted by an electrode on an otherwise insulating surface. Individual tubes in the bundle are clearly visible. The electrode is grounded, and the AFM tip biased at $-1$ V. Top inset: Topographic AFM image showing the bundle and the gold lead. Lower inset: Simultaneously acquired current image. Scan size in the inset is 850 nm.}
\label{fig1}
\end{figure}

\begin{figure}
\caption{(color) (a) Topographic image of a 8 nm diameter SWNT bundle contacted by two electrical leads. The top electrode is grounded, while the lower electrode is biased at $0.6$ V. (b) Image of the tip-sample current with the tip biased at 0.45 V. When the local voltage in the circuit is greater than the tip voltage, current flows from the sample to the tip (blue); when the tip voltage is larger, current flows from the tip to the sample (red). Coexistence of red and blue regions along the same bundle cross-section indicates that tubes within the bundle have weak electronic coupling.} 
\label{fig2}
\end{figure}

\begin{figure}
\caption{(color) (a) Topographic image of a SWNT bundle whose diameter varies from 1.5 - 8 nm contacted by two electrodes. (b) Higher magnification image of the boxed region in (a). The arrow marks a ``bottleneck" where the bundle narrows to $1.7$ nm, about the height of a single tube. (c) Image of the transport current as a function of tip (gate) position with $-2$ V applied to the tip. The bias voltage applied to the circuit is $0.2$ V. Contour lines superimposed on the image show that the local gating enhances the current significantly only in the vicinity of the bottleneck. (d) Transport current image recorded with a tip voltage of +1.5 V and 1 V bias voltage across the circuit. Gating action is again localized to the bottleneck region, but it shows an unusual ``p-n" structure.} \label{fig3}
\end{figure}

\end{document}